\documentstyle[aps,version2,preprint]{revtex}    
\begin{document}

\draft

\begin{title}
One-Particle Spectral Properties of Mott-Hubbard Insulators
\end{title}

\author{J. M. P. Carmelo$^{1,2}$, J. M. E. Guerra$^{2}$,\\
J. M. B. Lopes dos Santos$^{3}$,
and A. H. Castro Neto$^{4}$}
\begin{instit}
$^{1}$ N.O.R.D.I.T.A., Blegdamsvej 17, DK-2100 Copenhagen \O, Denmark
\end{instit}
\begin{instit}
$^{2}$ Department of Physics, University of \'Evora,
Apartado 94, P-7001 \'Evora Codex, Portugal
\end{instit}
\begin{instit}
$^{3}$ CFP and Departamento de F\'{\i}sica da Faculdade de
  Ci\^encias da Universidade do Porto\\
  Rua do Campo Alegre 687, P-4169-007 Porto, Portugal
\end{instit}
\begin{instit}
$^{4}$ Department of Physics, University of California,
Riverside, California 92521
\end{instit}
\receipt{April 1999}

\begin{abstract}
We use an exact {\it holon} and {\it spinon} Landau-liquid
functional which describes the holon - spinon interactions
at all scattering orders, to study correlation functions
of integrable multicomponent many-particle problems
showing both linear and non-linear energy bands. We consider
specific cases when the dominant non-linear band terms are
quadratic and apply our results to the evaluation of
the 1D Hubbard model correlation functions beyond
conformal-field theory.
\end{abstract}
\renewcommand{\baselinestretch}{1.656}   


\narrowtext

Recent experimental investigations on the one-particle spectral
properties of one-dimensional (1D) insulators \cite{Kobayashi}
have confirmed that the low-energy physics of these materials is
dominated by {\it holons} and {\it spinons} and that the conceptual
importance of these excitations is the same as that of other
elementary particles and quasiparticles in solids such as phonons
and magnons. The discovery of these elementary excitations followed
the exact diagonalization of the 1D Hubbard model \cite{Lieb},
whose Bethe-ansatz (BA) solution was obtained in terms of two coupled
equations, associated with charge and spin modes. However, that
solution does not provide direct information on matrix elements,
and previous theoretical studies on the one-particle spectral properties
of this model were mostly numerical \cite{Hanke,Penc}. Analytical studies
on the problem either referred to the limit of on-site Coulomb interaction
$U\rightarrow\infty$, when the charge - spin separation occurs at all
energies, or to the metallic phase away from half filling, where
conformal-field theory (CFT) has provided important information on
correlation functions \cite{Penc,Frahm,Carmelo96a}. Unfortunately,
that theory does not allow the evaluation of one-particle correlation
functions for the half-filling Mott insulator and the study of these
functions at finite values of $U$ remains an open problem of great
physical
interest.

In this Letter we combine a Landau-liquid functional which
describes exactly the holon and spinon interactions at all
scattering orders with new detected symmetries, to derive
asymptotic expressions for the 1D Hubbard model one-particle
correlation functions both at half filling and maximum
magnetization. Our critical theory refers to all integrable
multicomponent quantum liquids whose low-energy
physics is dominated by elementary excitations with both
nonlinear and linear energy band terms. We study the case
of quadratic bands in detail and also consider the general
case of nonlinear band edges. Our results reveal surprising
invariances of great importance for the theoretical
understanding and description of the unusual spectral
properties detected in the new low-dimensional materials.

We consider the 1D Hubbard model in a magnetic field $H$ and
chemical potential $\mu$ which describes $N=N_{\uparrow}+N_{\downarrow}$
interacting electrons and can be written as
$\hat{H}=\hat{T}+U[\hat{D}-{1\over 2}\hat{N}]
+\sum_{\alpha= c,s}\mu^h_{\alpha }2{\hat{S}}^{\alpha}_z$
where $\hat{T}=-t\sum_{j,\sigma}[c_{j\sigma}^{\dag}c_{j+1\sigma} + h. c.]$
is the ``kinetic energy'', $\hat{D} = \sum_{j}
\hat{n}_{j,\uparrow}\hat{n}_{j,\downarrow}$
measures the number of doubly occupied sites,
$\mu^h_c=\mu$, $\mu^h_s=\mu_0 H$, $\mu_0$ is the Bohr magneton,
${\hat{S }}^c_z=-{1\over 2}[N_a-\hat{N}]$, ${\hat{S }}^s_z=
-{1\over 2}[{\hat{N}}_{\uparrow}-{\hat{N}}_{\downarrow}]$,
${\hat{N}}=\sum_{\sigma}\hat{N}_{\sigma}$,
${\hat{N}}_{\sigma}=\sum_{j}\hat{n}_{j,\sigma}$,
$\hat{n}_{j,\sigma}=c_{j\sigma }^{\dagger }c_{j\sigma }$,
$c_{j\sigma}^{\dagger}$ and $c_{j\sigma}$ are electron
operators of spin projection $\sigma $ at site $j=1,...,N_a$,
and $t$ is the transfer integral. We choose a density
$n=n_{\uparrow}+n_{\downarrow}=N/N_a$ and a spin density
$m=n_{\uparrow}-n_{\downarrow}$ in the intervals $0\leq n\leq 1$
and $0\leq m\leq n$, respectively, where
$n_{\sigma}=N_{\sigma}/N_a$ and consider
$k_{F\sigma }=\pi n_{\sigma}$ and $k_F=\pi n/2$. We use units
such that $a=\hbar=1$, where $a$ is the lattice constant.

It is convenient to use the operational representation for
holon and spinon excitations introduced in Ref. \cite{Carmelo96}
which is exact and refers directly to the BA solution.
In this representation the holons and spinons are the {\it holes}
of the $c$ and $s$ anticommuting pseudoparticles \cite{Carmelo94},
respectively. It should be emphasized that this connection is only
true in the limit of zero magnetic field and that the concept of
$c$ and $s$ pseudoparticle is more general than that of holon and
spinon and refers also to finite values of $H$. In this basis the
Slater-determinant levels of $\alpha$ pseudoparticles, with $\alpha
=c,s$, are exact energy eigenstates of the many-electron problem
and the total momentum is additive in the pseudoparticle momenta $q$.
On the other hand, the Hamiltonian contains pseudoparticle interaction
terms, yet integrability implies that all the interactions are of
zero-momentum forward-scattering type. The exact many-electron ground
state
is in this representation described by the $\alpha$-pseudoparticle
distribution, $N_{\alpha}^0(q)=\Theta (q^{(+1)}_{F\alpha}-q)$ and
$N_{\alpha}^0(q)=\Theta (q-q^{(-1)}_{F\alpha})$ with
$0<q<q^{(+1)}_{\alpha}$ and $q^{(-1)}_{\alpha}<q<0$, respectively,
where $q^{(\pm 1)}_{F\alpha}=\pm q_{F\alpha}+O(1/N_a)$ and
$q_{F\alpha}=\pi N_{\alpha}/N_a$. Here $N_c=N$ and $N_s=N_{\downarrow}$
and hence $q_{Fc}=2k_F$ and $q_{Fs}=k_{F\downarrow}$.
The pseudo-Brillouin zones are $q^{(\pm 1)}_{\alpha}=
\pm q_{\alpha}+O(1/N_a)$ and $q_{\alpha}=\pi N^*_{\alpha}/N_a$
where $N^*_c=N_a$ and $N^*_s=N_{\uparrow}$ and hence
$q_{c}=\pi $ and $q_{s}=k_{F\uparrow}$. Note that at half filling,
$N^h_c=N^*_c-N_c=0$ (zero magnetization, $N^h_s=N^*_s-N_s=0$)
and the ground state has no holons (spinons). Importantly,
$N_{\alpha}$ and $N^*_{\alpha}$ are always conserving numbers.
The eigenstate energy expressions only involves
the momentum distribution $N_{\alpha}(q)$ and for states
differing from the ground state by a small density of
excited $\alpha $ pseudoparticles the excitation-energy spectrum
can be expanded in the momentum deviations,
$\Delta N_{\alpha}(q)=N_{\alpha}(q)-N_{\alpha}^0(q)$,
what leads to a Landau functional \cite{Carmelo94}, $\Delta E_L =
\sum_{i=1}^{\infty}\Delta E_i$, which up to third order
in $i$ reads

\begin{eqnarray}
\Delta E_L & = & \sum_{\alpha,q}\Delta N_{\alpha}(q)
\epsilon_{\alpha} (q) + {1\over L}
\sum_{\alpha ,\alpha'}\sum_{q,q'}\Delta N_{\alpha}(q)
\Delta N_{\alpha'}(q'){1\over 2}\,f_{\alpha,\alpha'} (q,q') +
\nonumber \\
& + & {1\over L^2}\sum_{\alpha,\alpha',\alpha''}\sum_{q,q',q''}
\Delta N_{\alpha}(q)\Delta N_{\alpha'}(q')
\Delta N_{\alpha''}(q'')
{1\over 6}\, g_{\alpha,\alpha',\alpha''} (q,q',q'') + h.o. \, .
\label{EL}
\end{eqnarray}
Importantly, the pseudoparticles do not map onto electrons as $U$
is adiabatically turned off.

All the coefficients of the functional (\ref{EL}) can be exactly
derived from the BA equations \cite{Carmelo94}, yet for orders
$i>2$ the calculations become extremely lengthy. So far
only the $i=1$ band $\epsilon_{\alpha} (q)$ and $i=2$
$f_{\alpha,\alpha'}$ function were derived \cite{Carmelo94}.
>From the band expressions the velocity $v_{\alpha}(q)=d\epsilon_{\alpha}
(q)/dq$ and the function $a_{\alpha}(q)=d v_{\alpha}(q)/dq$ follow.
The $f_{\alpha,\alpha'}$ function expression only involves the
velocity and the two-pseudoparticle phase shift $\Phi_{\alpha,\alpha'}
(q,q')$ \cite{Carmelo94}. This phase shift plays a central
role in our critical theory -- it is an interaction-dependent
parameter associated with the zero-momentum forward-scattering
collision of the $\alpha$ and $\alpha'$ pseudoparticles of momentum
$q$ and $q'$, respectively. For instance, the phase-shift combinations
$\xi_{\alpha\alpha'}^i=\delta_{\alpha,\alpha'}+\sum_{j=\pm 1}(j)^i
\Phi_{\alpha\alpha'} (q_{F\alpha},jq_{F\alpha'})$ with $i=0,1$
fully determined the $\alpha $ anomalous dimensions of the generalized
critical theory, as we discuss below.

When both the densities of $\alpha $ pseudoparticles, $n_{\alpha}$,
and of $\alpha $ pseudoholes, $n^h_{\alpha}$, are finite the dominant
$\epsilon_{\alpha} (q)$ band terms are linear in the
momentum for low-energy excitations. We emphasize that in this
case only the terms of order $i=1$ and $i=2$ of the functional (\ref{EL})
are relevant for the low-energy physics. Importantly,
replacing in expression (\ref{EL}) the bands and $f$ functions by the
corresponding linearized bands and values of the $f$ function at the
pseudo-Fermi momenta, using a general deviation expression describing
all low-energy processes, and performing the $q$ summations
leads to the CFT spectrum derived in Ref.
\cite{Frahm} by a finite-size analysis. This reveals that the
pseudoparticle basis is the natural representation for the critical-point
operator algebra. Moreover, for these densities CFT
provides asymptotic correlation functions expressions \cite{Frahm}.

We emphasize that the functional (\ref{EL}) is universal for a wide
class of integrable multicomponent models, with arbitrary number
of $\alpha $ pseudoparticle branches. It occurs often in these
systems that as the density $n_{\alpha}$ or $n^h_{\alpha}$ becomes
small the dominant low-energy $\alpha$ band term is non linear and
of the general form

\begin{equation}
\epsilon_{\alpha} (q)\approx\pm ||q|-Q_{\alpha}|^j
/j!{\cal{M}}_{\alpha}
\label{nlb}
\end{equation}
where $Q_{\alpha}=0$ when $n_{\alpha}\rightarrow 0$
and $Q_{\alpha}=q_{\alpha}$ when $n^h_{\alpha}\rightarrow 0$,
${\cal{M}}_{\alpha}$ is a positive generalized {\it mass},
and $j$ is a positive integer, $j=1,2,3,4,....,\infty$.
Below we use the notation $\{\alpha,j\}$ for a $\alpha$ band
of type $j$. In the present model there occur band changes
$\{c,1\}\rightarrow \{c,2\}$ as $n_c\rightarrow 0$ and
$n^h_c\rightarrow 0$, $\{s,1\}\rightarrow \{s,2\}$
as $n_s\rightarrow 0$, and $\{s,1\}\rightarrow\{s,\infty\}$
as $U\rightarrow\infty$ ($\epsilon_{s} (q)=0$ implies
$\{s,\infty\}$). A central point is that for each
$\{\alpha ,j\}$ branch the low-energy critical-theory spectrum has
contributions from terms up to order $i=j+1$ of the functional
$(\ref{EL})$.

In this Letter we consider explicitly the low-energy critical
theory associated with the case when some of the $\alpha $ branches
have $\{\alpha ,2\}$ bands and for all the remaining branches
(whose minimal number is one) the band is of $\{\alpha ,1\}$ type.
This requires the use of terms up to the order $i=3$ in the
functional (\ref{EL}). Lengthy calculations involving substitution
of the third-order deviation expansions in the BA equations
of Ref. \cite{Carmelo94} lead to $g_{\alpha,\alpha',\alpha''}
(q,q',q'')={\breve{g}}_{\alpha,\alpha',\alpha''}(q,q',q'')
+{\breve{g}}_{\alpha,\alpha',\alpha''}(q',q'',q)+{\breve{g}}(q''
,q,q')$ where

\begin{eqnarray}
{\breve{g}}_{\alpha,\alpha',\alpha''}(q,q',q'') & = & 4(\pi)^2\{
a_{\alpha} (q) \Phi_{\alpha,\alpha'} (q,q')\Phi_{\alpha,\alpha''} (q,q'')
\nonumber\\
& + & \sum_{j=\pm 1}\sum_{\alpha'''}j {a_{\alpha'''}\over 3}
\Phi_{\alpha''',\alpha}(jq_{F\alpha'''},q)
\Phi_{\alpha''',\alpha'} (jq_{F\alpha'''},q')
\Phi_{\alpha''',\alpha''} (jq_{F\alpha'''},q'')
\nonumber \\
& + & v_{\alpha}(q)[\Phi_{\alpha,\alpha''} (q,q'')
{d\over dq}\Phi_{\alpha ,\alpha'} (q,q')
+ {\cal{A}}_{\alpha,\alpha',\alpha''} (q,q',q'')]
\nonumber\\
& + & \sum_{j=\pm 1}\sum_{\alpha'''}v_{\alpha'''}
\Phi_{\alpha''',\alpha} (jq_{F\alpha'''},q)
{\cal{A}}_{\alpha''',\alpha',\alpha''} (jq_{F\alpha'''},q',q'')\} \, ,
\label{g}
\end{eqnarray}

\begin{eqnarray}
{\cal{A}}_{\alpha,\alpha',\alpha''} (q,q',q'') & = &
\Phi_{\alpha ,\alpha'} (q,q')
{d\over dq}\Phi_{\alpha ,\alpha''} (q,q'')
\nonumber \\
& + & 2\pi\rho_{\alpha}\left(r^{(0)}_{\alpha}(q)\right){d\over dq}
\{[{\Phi_{\alpha',\alpha} (q',q)\Phi_{\alpha',\alpha''} (q',q'')
\over 2\pi\rho_{\alpha'} \left(r^{(0)}_{\alpha'}(q')\right)} +
{\Phi_{\alpha'',\alpha} (q'',q)\Phi_{\alpha'',\alpha'} (q'',q')
\over 2\pi\rho_{\alpha ''} \left(r^{(0)}_{\alpha''}(q'')\right)}]
\nonumber \\
& + & \sum_{j=\pm 1}\sum_{\alpha'''}
j [{\Phi_{\alpha''',\alpha} (jq_{F\alpha'''},q)
\Phi_{\alpha''',\alpha'} (jq_{F\alpha'''},q')
\Phi_{\alpha''',\alpha''} (jq_{F\alpha'''},q'')
\over 2\pi\rho_{\alpha '''} \left(r_{\alpha'''}\right)}]\} \, ,
\label{A}
\end{eqnarray}
$v_{\alpha }\equiv v_{\alpha}(q_{F\alpha })$,
$a_{\alpha }\equiv a_{\alpha}(q_{F\alpha })$,
the function $2\pi\rho_{\alpha}\left(r\right)$ is
such that $1/2\pi\rho_{\alpha}\left(r^{(0)}_{\alpha}(q)\right)=
dr^{(0)}_{\alpha}(q) /dq$, $r^{(0)}_{c}(q)=K^{(0)}(q)$,
$r^{(0)}_{s}(q)=S^{(0)}(q)$, $K^{(0)}(q)$ and
$S^{(0)}(q)$ are the rapidity functions
obtained by solving the integral equations $(A1)$
and $(A2)$ of the second paper of Ref. \cite{Carmelo94} for
the ground-state distribution $N^0(q)$, and
$r_{\alpha}=r^{(0)}_{\alpha}(q_{F\alpha})$.

The Landau character of the spectrum (\ref{EL}) implies
that {\it the only} small parameter which defines the
range of validity of the corresponding low-energy critical
theory is the density of excited pseudoparticles. The
associate low-energy processes can produce changes in the
numbers $N_{\alpha \iota}$ of right ($\iota =+1$) and left
($\iota=-1$) pseudoparticles (with $\iota =sgn (q)1=\pm 1$),
which are conserved independently. Equivalently, they can
produce changes in the charge and current numbers $N_{\alpha
}=\sum_{\iota} N_{\alpha \iota}$ and $J_{\alpha }=(1/2)
\sum_{\iota}\iota N_{\alpha \iota}$, respectively. The
above processes can  be classified into three types, (i)
ground-state -  ground-state transitions associated with
variations $\Delta N_{\alpha }=N_{\alpha }-N_{\alpha }^0$
which only change $q_{F\alpha }$, (ii) finite  momentum
$K=\sum_{\alpha}{\cal{D}}_{\alpha }2q_{F\alpha }$ processes
associated with variations ${\cal{D}}_{\alpha }=J_{\alpha
}-J_{\alpha }^0$, and (iii) a number $N^{\alpha\iota}_{ph}=
0,1,2,...$ of elementary particle - hole processes around
the Fermi point $q^{(\iota)}_{F\alpha}$. In the absence of
transitions (i) we define $q_{F\alpha }$ relatively to the
initial ground state and otherwise relatively to the final
ground state. Independently of the form of the band
$\epsilon_{\alpha } (q)$, at low energy the above processes lead
to a momentum of the form $K=k_0+\sum_{\alpha}\Delta P_{\alpha}$
where $\Delta P_{\alpha}={2\pi\over N_a}[\Delta N_{\alpha }
{\cal{D}}_{\alpha } + \sum_{\iota}\iota N_{\alpha\iota}^{ph}]$
and in the present case, $k_0=\sum_{\alpha }{\cal{D}}_{\alpha }
2q_{F\alpha}$. Let us denote by $\sum_{\bar{\alpha}}$,
$\sum_{\breve{\alpha}}$, and $\sum_{\alpha}$ the summations
over linear branches, quadratic (or other $j>2$ non-linear)
branches, and all types of branches, respectively.
The $\{\alpha ,2\}$ band reads $\epsilon_{\alpha} (q)
=\pm ||q|-Q_{\alpha}|^j/2m^*_{\alpha}$, with ${\cal{M}}_{\alpha}
\equiv m^*_{\alpha}$ and $m_{\alpha}^*=1/a_{\alpha} (0)$
and $m_{\alpha}^*=1/a_{\alpha}(q_{\alpha})$ for pseudoparticles
and pseudoholes, respectively. These masses play the same role
as the effective mass of Ref. \cite{McGuire}. Introducing in Eqs.
(\ref{EL})-(\ref{A}) the $\{\alpha ,j\}$ band expressions
(with $j=1$ or $j=2$) and the low-energy deviations associated
with all low-energy processes and performing the $q$ summations
leads to the critical spectrum

\begin{equation}
\Delta E = {2\pi\over L}[\sum_{\bar{\alpha}}v_{\alpha }
\Delta^E_{\alpha} + \sum_{\breve{\alpha}}{\breve{v}}_{\alpha}
({\cal{N}}^1_{\alpha})^2] + \sum_{\breve{\alpha}}\{
({\breve{v}}_{\alpha})^3{L\over 6\pi}(m_{\alpha}^*)^2
+ ({2\pi\over L})^2{{\cal{E}}_{\alpha}\over m_{\alpha}^*} +
{2\pi\over L}{\breve{v}}_{\alpha} N_{\alpha}^{ph}\} \, ,
\label{MDE}
\end{equation}
where $\Delta^E_{\alpha}=\sum_{i=0,1}({\cal{N}}^i_{\alpha})^2
+N_{\alpha}^{ph}$, ${\cal{N}}^0_{\alpha}=\sum_{\alpha'}
\xi_{\alpha\alpha'}^0{\Delta N_{\alpha'}\over 2}$,
${\cal{N}}^1_{\alpha}=\sum_{\alpha'}\xi_{\alpha\alpha'}^1
{\cal{D}}_{\alpha'}$, ${\breve{v}}_{\alpha}=
v_{\alpha}\xi_{\alpha}^0$
with ${\breve{v}}_{\alpha}=
q_{F\alpha}/{\breve{m}}_{\alpha}$ and
$v_{\alpha}=q_{F\alpha}/m_{\alpha}^*$
(or ${\breve{v}}_{\alpha}=
[q_{\alpha}-q_{F\alpha}]/{\breve{m}}_{\alpha}$ and
$v_{\alpha}=[q_{\alpha}-q_{F\alpha}]/m_{\alpha}^*$),
$\xi_{\alpha}^0 = {m_{\alpha}^*\over {\breve{m}}_{\alpha}}
= {{\breve{v}}_{\alpha}\over v_{\alpha}} =
{{\cal{N}}^0_{\alpha}\over \Delta N_{\alpha}}$,
and ${\cal{E}}_{\alpha}$ is independent of
${\breve{v}}_{\alpha}$ and contains no pseudoparticle
interaction parameters. Note that
$q_{F\alpha}=\pi\Delta N_{\alpha}/N_a$
(or $q_{\alpha}-q_{F\alpha}=-\pi\Delta N_{\alpha}/N_a$)
in the limit when $n_{\alpha}=0$ (or $n^h_{\alpha}$=0) for the initial
ground state.

When all $\alpha $ bands are linear, the critical-theory expressions
involve the anomalous dimensions, $2\Delta_{\alpha}^{\iota}=
\sum_{i=0,1}({\cal{N}}^i_{\alpha})^2+\iota \Delta N_{\alpha}
{\cal{D}}_{\alpha}+2N_{\alpha,\iota}^{ph}$ (equal to
$(\sum_{i=0,1}\iota^i{\cal{N}}^i_{\alpha})^2+
2N_{\alpha,\iota}^{ph}$ when there is dual symmetry) and
the asymptotics of correlation functions is given by CFT
\cite{Frahm} and reads $\chi_{\vartheta}(x,t) \propto
{\prod_{\alpha,\iota}[e^{-ik_0 x}/
(x-\iota v_{\alpha}t)^{2\Delta^{\iota}_{\alpha}}}]$.
On the other hand, the critical theory associated
with the low-energy spectrum
(\ref{MDE}) is characterized by two remarkable properties: (I)
in spite of its non-linear velocity terms, {\it the same} above
anomalous dimensions are obtained from the following
equation, $2\Delta_{\alpha}^{\iota}=
{L\over 2\pi}[\sum_{\bar{\alpha}}d\Delta E/dv_{\alpha} +
\sum_{\breve{\alpha}}d\Delta E/d{\breve{v}}_{\alpha}
+\iota\Delta P_{\alpha}]$. Since their expressions are the
same as in the linear regime, their values are given by taking
the corresponding limits of the densities in the general
expressions -- in the present model the values of the dimensions
$2\Delta_{c}^{\iota}$ (or $2\Delta_{s}^{\iota}$) are in the quadratic
case given by the limit $n\rightarrow 1$ (or $m\rightarrow n$) of the
corresponding dimensions of the linear regime; and
(II) as $n_{\alpha}\rightarrow 0$ or $n^h_{\alpha}\rightarrow 0$
there is an {\it adiabatic} symmetry \cite{note} which implies that
$\xi_{\alpha\alpha'}^0 \, ,\xi_{\alpha'\alpha}^1 \rightarrow
\delta_{\alpha,\alpha'}$ in these limits. (However, if the $\alpha '$
band is linear, $\xi_{\alpha'\alpha}^0$ and $\xi_{\alpha\alpha'}^1$
remain in general interaction dependent.) This symmetry implies that
$\xi_{\alpha}^0\rightarrow 1$, {\it i.e.}
the pseudoparticle interactions which renormalize the quadratic
mass $m_{\alpha}^*$ and associate velocity $v_{\alpha}$ vanish
(and ${\breve{m}}_{\alpha}=m_{\alpha}^*$). For instance,
in the present model
$\xi_{c\alpha}^0\, , \xi_{\alpha c}^1\rightarrow \delta_{c,\alpha}$
as $n\rightarrow 1$ ({\it i.e.}, $n^h_c\rightarrow 0$) and
$\xi_{s\alpha}^0\, , \xi_{\alpha s}^1\rightarrow \delta_{s,\alpha}$
as $m\rightarrow n$ ({\it i.e.}, $n_s\rightarrow 0$).

These symmetries  are observed in all multicomponent integrable models
and imply that the terms of the critical spectrum (\ref{MDE}) which
contain no linear velocity terms are of non-interacting pseudoparticle
character and only the linear terms in the velocities are interacting.
Following this remarkable property and associate above symmetries, we
find the following general expressions for the asymptotics of
correlation functions

\begin{eqnarray}
\chi_{\vartheta } (x,t) & \propto &
\prod_{\breve{\alpha},\bar{\alpha} ',\iota'}e^{-ik_0 x}/
[(x-\iota' v_{\alpha '}t)^{2\Delta^{\iota'}_{\alpha'}}
\,(x)^{\sum_{\iota}2\Delta^{\iota}_{\alpha}}] \, ,
\hspace{0.5cm} x/\sqrt{2t/m_{\alpha }^*}>> 1 \, ,
\hspace{0.5cm} \alpha\in\breve{\alpha}  \nonumber \\
& \propto &
{\prod_{\breve{\alpha},\bar{\alpha}',\iota'}e^{-ik_0 x}/
[(-\iota' v_{\alpha'}t)^{2\Delta^{\iota'}_{\alpha'}}
\,(\sqrt{2t/m_{\alpha }^*})^{\sum_{\iota}2\Delta^{\iota}_{\alpha}}}]
\, , \hspace{1cm} x/t = 0 \, .
\label{cf}
\end{eqnarray}
These $x$ and $t$ dependences can be understood in the following
way. In the limit of low energy each $\alpha $ excitation branch
corresponds to an independent momentum-energy tensor component and
to one independent Minkowski space with {\it light velocity}
$v_{\alpha}$. For bands with both finite $n_{\alpha}$ and
$n^h_{\alpha}$ densities the velocity $v_{\alpha}$
is also finite and the metric is Lorentzian. However, for
vanishing small values of $n_{\alpha}$ or $n^h_{\alpha}$ the
metric becomes Galilean. In the Lorentzian case the
asymptotic of correlation functions involves the variables
$(x\pm v_{\alpha}t)$ associated with Lorentzian
transformations. On the other hand, in the case of Galilean symmetry
space $x$ and time $t$ are transformed independently. This is
consistent with the two asymptotic-expression regimes involving
either $x$ or $t$.
Moreover, the only combination of the time $t$ and mass $m_{\alpha}^*$
with dimensions of $x$ is $const\times\sqrt{t/m_{\alpha}^*}$.
Note that the changes in the asymptotics only
concern the metric whereas the $\alpha$ anomalous dimensions,
whose values depend on $U/t$, $n$, and $m$, remain the
same. Both the CFT asymptotic correlation function expressions and
expressions (\ref{cf}) are limiting cases valid for finite and
vanishing, respectively, values of $n_{\alpha}$ or $n^h_{\alpha}$. As
$n_{\alpha}$ or $n^h_{\alpha}$ is gently increased, we come into a
small-density $\{\alpha, 2\}\rightarrow\{\alpha ,1\}$ band
transition regime which is not described by these asymptotic
expressions.

Our results can be generalized for low-energy $j=1$ and $j>1$
bands of general form (\ref{nlb}). However,
our theory {\it does not} describe the case when all bands
are of $j>1$ type. Although for $j>2$ we have not calculated
explicitly the coefficients of the functional (\ref{EL}) up
to order $i=j+1$, symmetry considerations lead to expression
(\ref{cf}) for $x/(t/{\cal{M}}_{\alpha})^{1/j}>> 1$,
whereas the $x/t=0$ expression becomes
$\chi_{\vartheta}(x,t)\propto {\prod_{\breve{\alpha},\bar{\alpha}',\iota'}
e^{-ik_0 x}/[(-\iota' v_{\alpha'}t)^{2\Delta^{\iota'}_{\alpha'}}\,
(t/{\cal{M}}_{\alpha})^{{\sum_{\iota}2\Delta^{\iota}_{\alpha}\over j}}}]$.
Here $j$ is meant to be a function of $\alpha$, {\it i.e.}
different $\alpha$ bands may have different $j>1$ values.
As we refer below, these expressions lead for
the Hubbard model $U\rightarrow\infty$ $\{c,1\}$ and $\{s,\infty\}$
bands to the same correlation function expressions
as Ref. \cite{Penc}.

Fourier transforms of the above asymptotic expansions provide
correlation function expressions for values of momentum
$k$ close to $k_0$ and low values of energy $\omega$
measured from the initial ground-state energy. For
instance, let $A_{\sigma}(k,\omega)$,
$N_{\sigma}(k)=\sum_{\omega} A_{\sigma}(k,\omega)$,
and $D_{\sigma}(\omega)=\sum_k A_{\vartheta}(k,\omega)$
be the one $\sigma $ electron (or hole) spectral function,
momentum distribution, and density of states
of the Hubbard chain. There are no previous
analytical results for these functions at finite values of $U$
and for (I) the half-filling Mott-Hubbard insulator ($n=1$) and (II) the
fully
polarized ferromagnetic ($n_{\downarrow}=0$ and $n_{\uparrow}=n=m$)
initial ground states. In case I the bands are of
$\{c,2\}$ and $\{s,1\}$ type. For $m=0$ and both
$n=1$ and small finite densities of holes $\delta $ and low negative
(positive) values of $\omega $ and/or values of $k$
close to $k_F$ our generalized theory leads to
to $A_{\sigma} (k_F,\omega)\propto |\omega|^{-7/8}$ and
$N_{\sigma}(k)\propto |k-k_F|^{1/8}$ for the particle
(hole) spectral function and momentum distribution.
(At $n=1$ we define the ground-state energy at the bottom and
the top of the Mott-Hubbard gap \cite{Lieb} for particles and
holes, respectively.) On the other hand, the particle (hole)
density of states is given by $D_{\sigma}(\omega)\propto
|\omega|^{-3/16}$ and $D_{\sigma}(\omega)\propto
|\omega |^{1/8}$ for $n=1$ and small finite
$\delta $ values, respectively. However, the latter
expression (also predicted by CFT) is
restricted to frequencies $|\omega |
<E_c=\delta ({2\pi\over L})^2 {5\over 32\, m_c^*}$.
In the limit of $n\rightarrow 1$ this domain shrinks
to a single point and the spectral function diverges as
$D(\omega)\propto |\omega |^{-3/16}$.
In case II the bands are of $\{c,1\}$ and $\{s,2\}$ type. We
consider creation of one $\downarrow$ electron and find for
both $m=n$ and for small finite densities $n_{\downarrow} $,
$A_{\downarrow} (k_{F\downarrow},\omega)\propto
|\omega|^{-1+{1\over 2}[1-\eta_0]^2}$ and
$N_{\downarrow}(k)\propto |k-k_{F\downarrow}|^{{1\over 2}[1-\eta_0]^2}$,
where $\eta_0=(2/\pi)\tan^{-1}([4t\sin (\pi n)]/U)$
and $n<1$, whereas the density of states is given by
$D_{\downarrow}(\omega)\propto
|\omega|^{-{1\over 2}+{1\over 2}[1-\eta_0]^2}$
and $D_{\downarrow}(\omega)\propto
|\omega|^{{1\over 2}[1-\eta_0]^2}$ for $m=n$ and small
finite values of $n_{\downarrow}$, respectively.
The latter expression is restricted to energies
$|\omega | <E_s=\delta ({2\pi\over L})^2 {1\over m_s^*}$.
Again, in the limit of $n_{\downarrow}\rightarrow 0$ this
domain shrinks to a point and the spectral function
behaves as $D_{\downarrow}(\omega)\propto
|\omega|^{-{1\over 2}+{1\over 2}[1-\eta_0]^2}$.
These $D_{\downarrow}(\omega)$ expressions
are not valid for $U\rightarrow\infty$
because then the bands are of $\{c,1\}$ and $\{s,\infty\}$
type and instead $D_{\downarrow}(\omega)\propto |\omega|^{-{1\over 2}}$.
While for $n<1$ all our expressions lead to the correct
results as $U\rightarrow 0$, the pseudoparticle theory
also describes correctly the $n=1$ Mott-Hubbard transition
at $U=0$ \cite{Lieb}, the $c$ band being of $\{c,2\}$ and
$\{c,1\}$ type for $U>0$ and at $U=0$, respectively.
Finally, concerning the comparison of our results with
previously obtained $m=0$ and $U\rightarrow\infty$ expressions,
while our theory does not apply to the $n=1$ case \cite{Carmelo96a}
of bands $\{c,2\}$ and $\{s,\infty\}$,
it provides expressions for the $\{c,1\}$ and $\{s,\infty\}$
case which corresponds to finite values of $n$
and $\delta$. Importantly, our general expressions
lead to the same results as Ref. \cite{Penc}, with the density of
states given by $D_{\sigma}(\omega)\propto
|\omega|^{-3/8}$ and $D_{\sigma}(\omega)\propto
|\omega |^{1/8}$ for $U\rightarrow\infty$ and
small finite values of $4t/U$, respectively.
Our results reveal that the same anomalous dimensions
control the low-energy critical theory independently
of the linear or nonlinear character of its elementary-excitation
energy bands and are expected to shed new light
on the unusual properties of quasi-1D materials.

We thank D. K. Campbell, A. Luther, and L. M. Martelo
for illuminating  discussions and the support of PRAXIS
under Grants No. 2/2.1/FIS/302/94 and BCC/16441/98. A. H. C. N.
acknowledges support from the Alfred P. Sloan Foundation and
the partial support provided by a CULAR research grant under the
auspices of the US Department of Energy.



\begin{references}
\bibitem[10]{note}
        In analogy with the adiabatic vanishing of the Fermi-liquid
        quasiparticle interactions, which here is achieved by
        adiabatically changing $n_{\alpha}$ or $n^h_{\alpha}$.
\end{references}
\end{document}